\documentclass[%
 reprint,
 amsmath,amssymb,superscriptaddress,
 aps,
]{revtex4-2}

\usepackage{graphicx}

\usepackage{bm}
\usepackage{braket}
\usepackage{amsmath}

\usepackage{float}
\usepackage[colorlinks=true,linkcolor=blue,citecolor=blue,urlcolor=blue]{hyperref}

\begin{document}

\title{Kibble-Zurek Mechanism in the Open Quantum Rabi Model}

\author{T. Pirozzi}
\affiliation{Dipartimento di Fisica “Ettore Pancini”, Università degli Studi di Napoli Federico II,
Complesso Universitario di Monte Sant’Angelo, Via Cintia, 80126 Napoli, Italy}

\author{G. Di Bello}
\affiliation{Dipartimento di Fisica “Ettore Pancini”, Università degli Studi di Napoli Federico II,
Complesso Universitario di Monte Sant’Angelo, Via Cintia, 80126 Napoli, Italy}
\affiliation{INFN, Sezione di Napoli - Complesso Universitario di Monte Sant’Angelo, Via Cintia, 80126 Napoli, Italy}

\author{V. Cataudella}
\affiliation{INFN, Sezione di Napoli - Complesso Universitario di Monte Sant’Angelo, Via Cintia, 80126 Napoli, Italy}
\affiliation{SPIN-CNR and Dipartimento di Fisica “Ettore Pancini” - Università di Napoli Federico II, 80126 Napoli, Italy}

\author{C. A. Perroni}
\affiliation{INFN, Sezione di Napoli - Complesso Universitario di Monte Sant’Angelo, Via Cintia, 80126 Napoli, Italy}
\affiliation{SPIN-CNR and Dipartimento di Fisica “Ettore Pancini” - Università di Napoli Federico II, 80126 Napoli, Italy}

\author{G. De Filippis}
\affiliation{INFN, Sezione di Napoli - Complesso Universitario di Monte Sant’Angelo, Via Cintia, 80126 Napoli, Italy}
\affiliation{SPIN-CNR and Dipartimento di Fisica “Ettore Pancini” - Università di Napoli Federico II, 80126 Napoli, Italy}

\date{\today} 
\begin{abstract}
The Kibble-Zurek mechanism provides a universal framework for predicting defect formation in non-equilibrium phase transitions. While Markovian dissipation typically degrades universal scaling, the impact of non-Markovian memory remains largely unexplored. We demonstrate that an Ohmic bath induces a Berezinskii-Kosterlitz-Thouless transition in the open quantum Rabi model. Using simulations based on Matrix Product States, we show that the excitation energy strictly follows universal Kibble-Zurek power-law scaling when evaluated at the freeze-out time. Crucially, we find that since the environment defines the universality class, dissipation does not inherently compete with adiabatic dynamics, in stark contrast to Markovian regimes. Our results establish the Kibble-Zurek mechanism as a robust witness of universality in open quantum systems, revealing that non-Markovian memory preserves the integrity of non-equilibrium scaling.
\end{abstract}
\maketitle

Understanding the non-equilibrium response of many-body quantum systems to time-dependent driving is a central objective of contemporary physics \cite{Kim2023,sperim}. This frontier is currently explored with unprecedented resolution in architectures such as superconducting circuits \cite{Acharya2023}, trapped ions \cite{Moses2023}, and Rydberg atom arrays \cite{Evered2023}, which provide a unique vantage point to probe quantum phase transitions (QPTs) \cite{sachdev2011} with a level of control previously beyond reach. These platforms enable the direct interrogation of many-body criticality, where quantum fluctuations drive an abrupt change in the nature of the ground state. In this context, a primary challenge lies in the preparation of highly correlated states by driving the system across a QPT, a process central to both quantum state engineering and adiabatic quantum computation. However, the very nature of criticality, marked by a vanishing energy gap, poses a formidable barrier to such adiabatic evolution. Near the critical point, the divergence of the relaxation time, or critical slowing down, inevitably forces a breakdown of adiabaticity for any finite-time protocol. This process leads to the production of excitations that not only limit the performance of quantum protocols but also serve as a universal dynamical signature of the underlying phase transition.

The Kibble-Zurek mechanism (KZM) has established itself as a cornerstone for understanding the non-equilibrium dynamics of systems traversing a quantum phase transition \cite{Kibble1976,Zurek1985}. By identifying a universal scaling law between the quench velocity and the production of excitations, the KZM serves as a powerful dynamical witness of the underlying universality class. Far from the transition, the dynamics remain nearly adiabatic, allowing the system to track its instantaneous ground state. However, as the critical point is approached, the relaxation time diverges. Once it becomes comparable to the quench timescale and eventually exceeds it, the adiabatic approximation is violated and a freeze-out occurs: the system effectively freezes, lacking the time to instantaneously respond to the quench. At this threshold, the correlation length ceases to grow and becomes frozen at a value dictated by the quench rate, leaving a persistent imprint on the final state. Since physical observables are determined by the correlation length at this point, they exhibit universal power-law scaling with respect to the quench velocity, revealing that the non-equilibrium dynamics are governed solely by the speed of the protocol and the universality class of the transition. The Kibble-Zurek mechanism has been experimentally verified across a remarkably diverse range of architectures, including liquid crystals \cite{LiquidCrystal}, superfluid helium \cite{Helium1,Helium2}, ultracold atomic gases \cite{UltracoldGas}, and ion crystals \cite{ioniccrystal}. While the mechanism is traditionally framed in terms of the growth of spatial correlation lengths, its validity extends far beyond systems with a well-defined spatial structure. In many-body systems lacking such structure, the standard interpretation based on spatial domains is no longer applicable; nevertheless, evidence suggests that the KZM framework remains remarkably robust \cite{PlenioPuebla-Rabi,Caneva2008}. In these cases, the universal scaling of physical observables, most notably the excitation energy, can be formally recovered through adiabatic perturbation theory \cite{Polkovnikov1}. This establishes the KZM as a witness of criticality even in zero-dimensional systems, demonstrating that non-equilibrium scaling laws can transcend the requirement of spatial dimensionality. 

Despite the robustness of the KZM in isolated settings, a significant conceptual hurdle persists: modern architectures are intrinsically coupled to their surroundings. This interaction transforms them into open quantum systems, where the interplay between critical dynamics and dissipation can fundamentally alter the nature of the transition, turning the quest for universal scaling into a battle against environmental noise. While the KZM is well-consolidated for closed systems, its extension to the open regime remains a frontier of intense debate. In the presence of Markovian dissipation, the KZM is often significantly compromised. Although a memoryless environment may not always change the universality class itself, it introduces an extrinsic dissipation rate that competes with both the quench velocity and the system's internal relaxation time \cite{AntiKZ1}. This competition typically leads to a breakdown of standard scaling in the slow-quench limit, where dissipation dominates the dynamics and prevents the system from accessing the purely adiabatic regime \cite{Ptimal}. In this regime, evidence suggests that bath-induced dissipation can generate excitations that counteract the adiabatic suppression of defects, leading to a departure from standard power laws often associated with anti-Kibble-Zurek scaling \cite{AntiKZ-plenio}. The resulting overabundance of excitations represents a major bottleneck for the performance of quantum simulators and adiabatic protocols. The situation becomes fundamentally more complex when considering non-Markovian dissipation. Unlike their Markovian counterparts, non-Markovian environments possess long-lived memory effects that can profoundly reshape the equilibrium properties of the system. Indeed, bath memory has been shown to shift critical points, change the order of transitions, or even fundamentally alter the universality class by inducing effective long-range interactions \cite{Nonmarkov-tfim,Bando2020,nonmarkov2}. Given these radical changes, one might plausibly expect a breakdown of the KZM, with dissipation acting as a source of additional excitations that overwhelm the universal scaling. Whether the KZM can still provide a robust description of the dynamics under these conditions remains a non-trivial and largely unexplored question.

In this Letter, we investigate the validity of the KZM in the Open Quantum Rabi Model (OQRM) coupled to a non-Markovian Ohmic bath, revealing a regime where universal scaling is remarkably preserved despite the presence of dissipation. This system serves as a uniquely compelling case study: the interaction with the environment fundamentally induces a Berezinskii-Kosterlitz-Thouless (BKT) phase transition \cite{kosterlitz1973,kosterlitz1981,bkt2006}, marking a stark departure from the second-order transition of the isolated model \cite{PlenioPuebla-Rabi}. This shift allows us to isolate the profound impact of non-Markovian memory effects on critical dynamics in a cornerstone model of quantum optics. To tackle the complex dynamics of this system, we employ state-of-the-art numerical techniques based on Matrix Product States(MPS) \cite{white1992dmrg,schollwock2011age}. Specifically, we use Density Matrix Renormalization Group (DMRG) for ground-state characterization and the Time-Dependent Variational Principle (TDVP) for time evolution \cite{paeckel2019time,haegeman2016unifying}. Simulating this model is notoriously difficult \cite{chin2010exact,zaletel2015time}: near the quantum critical point, the divergence of bosonic excitations requires the inclusion of a vast number of environmental degrees of freedom to capture the bath-induced criticality and memory effects. Our results provide the first dynamical evidence of BKT universality in this model, obtained by analyzing the relaxation dynamics following an external excitation. By subsequently performing quench protocols across a wide range of velocities, we demonstrate that the excitation energy follows a clear power-law scaling, as predicted by the KZM. Notably, this scaling behavior can be analytically understood by mapping the critical dynamics onto an effective Landau-Zener problem \cite{Damski2005}. This approach reveals that, in the slow-quench limit, the non-equilibrium response is governed by the non-adiabatic crossing of a renormalized two-level system, providing a transparent physical interpretation of the observed exponents. Crucially, our findings show that, unlike the defect overabundance typical of Markovian regimes, non-Markovian memory preserves the integrity of universal scaling even in the absence of spatial structure. Since the primary effect of the non-Markovian bath is to redefine the universality class rather than acting as a source of extrinsic noise, dissipation does not inherently compete
with adiabatic dynamics. Instead, it establishes the very universality to which the scaling belongs. Our work thus establishes the KZM as a robust witness of criticality in the open quantum regime, even when the environment itself dictates the underlying physics. 

\textit{The model: QPT evidence.} The system is described by the Hamiltonian $H = H_{\text{Q-O}} + H_I$. The first term, 
\begin{equation}
    H_{\text{Q-O}} = -\frac{\Delta}{2}\sigma_x + \omega_0 a^\dagger a + g \sigma_z(a^\dagger + a),
\end{equation}
represents the qubit-oscillator system with tunneling energy $\Delta$, resonator frequency $\omega_0$, and coupling $g$. The second term, $H_I = \sum_{i} [p_i^2/2M_i + k_i(x-x_i)^2/2]$, accounts for a bath of harmonic oscillators coupled to the resonator coordinate $x = \sqrt{1/2m\omega_0}(a+a^\dagger)$. The environment is characterized by an Ohmic spectral density $J(\omega) = \alpha \omega \Theta(\omega_c - \omega)$, where $\alpha$ is the dissipation strength and $\omega_c$ the cutoff frequency. We set $\hbar=1$.
 
As shown in Ref.~\cite{zueco,signature}, exact diagonalization of the bosonic sector maps the system onto an effective spin-boson model with a structured bath. Even in the low-dissipation regime ($\alpha \ll 1$), non-Markovian memory effects drive a BKT phase transition at a critical coupling determined by the condition $4g^2 \alpha \sim \omega_0^2$. This critical point separates a delocalized phase, characterized by spin tunneling between the up and down states, from a localized phase, where the tunneling amplitude is renormalized to zero when $\omega_c \rightarrow \infty$ \cite{lehur-entangl,LeHur2008}. Integrating out the bath degrees of freedom yields an effective Euclidean action~\cite{effective-Feynman,weiss,signature}. This maps the problem onto a one-dimensional classical spin chain with long-range ferromagnetic interactions scaling as $1/\tau^2$. This system exhibits a BKT transition, where the order parameter is defined by the squared magnetization fluctuations~\cite{signature},
   $ M^2 = \lim_{\beta \to \infty} \frac{1}{\beta} \int_0^\beta d\tau \langle \sigma_z(\tau) \sigma_z(0) \rangle.$ In the following, we focus on the parameter set $\alpha = 0.2$, $\omega_c = 10 \Delta$, and $\omega_0 = 0.75 \Delta$. A precise estimate of the critical coupling, obtained via world-line quantum Monte Carlo simulations and finite-size scaling analysis \cite{minnhagen,minnhagen2}, gives $g_c/\Delta = 0.9165$. 

\begin{figure}[H]
    \centering
    \includegraphics[width=\linewidth]{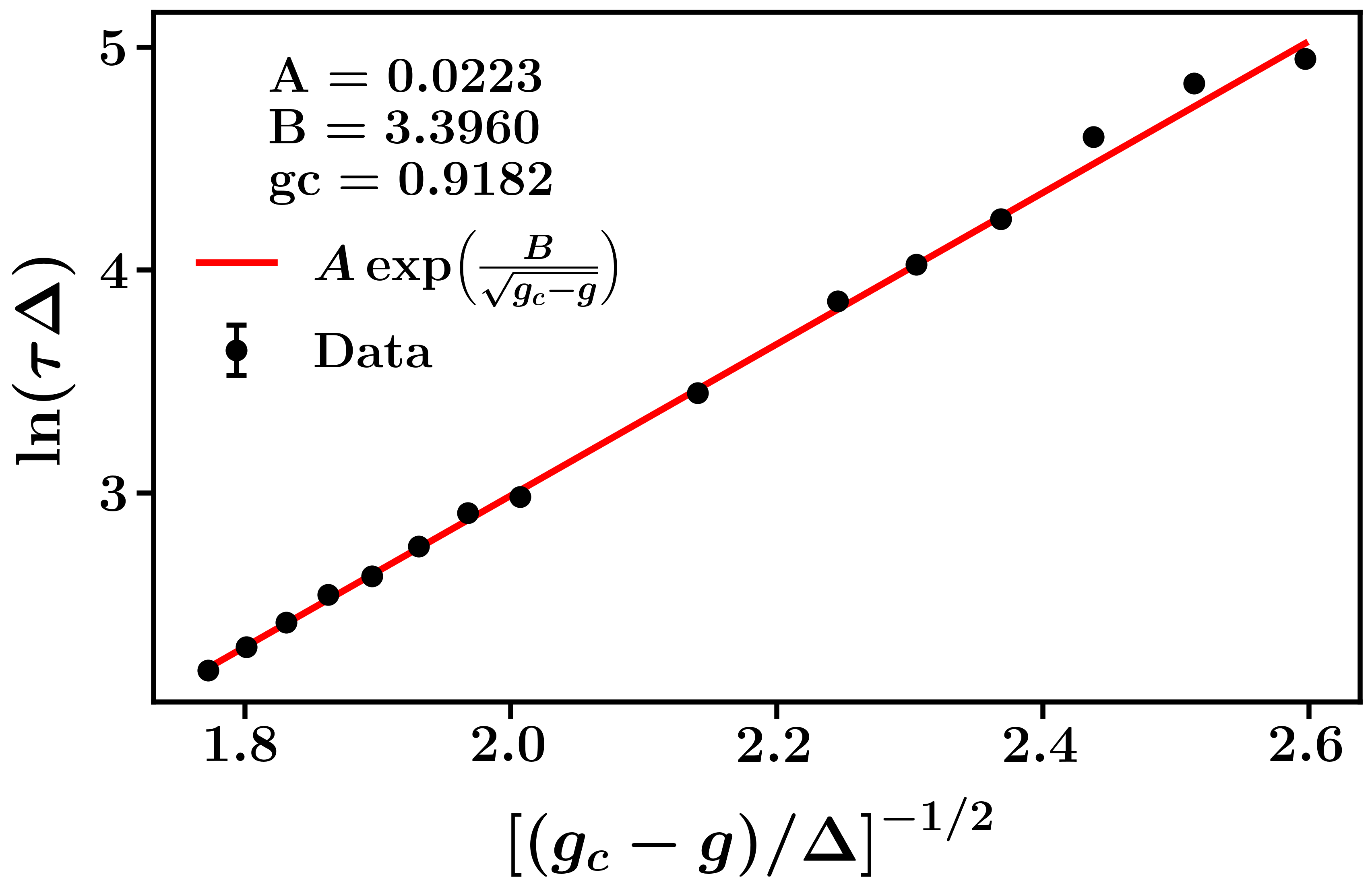}
    \caption{Relaxation time $\tau$ vs $g/\Delta$: numerical data (black dots) and fit to the BKT scaling $\tau \propto \exp[B/\sqrt{g_c - g}]$ (solid red line), yielding $g_c/\Delta \simeq 0.918$.}
    \label{fig:relaxation}
\end{figure}

\textit{Relaxation dynamics.} To characterize the critical properties of the OQRM, we first investigate the relaxation dynamics of the longitudinal spin magnetization. Following a linear response protocol \cite{Kubo1957}, the system is prepared in the ground state of $H + \epsilon \sigma_z$ ($\epsilon \ll \Delta$) and, at $t=0$, the longitudinal perturbation is abruptly removed. The subsequent evolution toward equilibrium is monitored through the normalized relaxation function 
\begin{equation}
    \Sigma_z(t) = \langle \sigma_z(t) \rangle / \langle \sigma_z(0) \rangle
\end{equation}
We simulated the relaxation dynamics using the TDVP  within the ITensor library \cite{itensor}, employing a star-geometry discretization to resolve the Ohmic spectral density $J(\omega)$. Although the resulting long-range interactions make the simulations computationally highly demanding, the MPS representation efficiently encodes this complexity, allowing us to capture the full non-Markovian back-action of the environment on the system \cite{MPSlongrange,GDFdynamics,DiBello2024,Parlato2025}. The relaxation time $\tau$ is extracted from the long-time behavior of $\Sigma_z(t)$, where a marked critical slowing down is observed as $g$ approaches the critical value. As shown in Fig.~\ref{fig:relaxation}, the dependence of $\tau$ on the coupling is in excellent agreement with the BKT scaling form \cite{kosterlitz1973,scalingbkt-kosterlitz1974}:
\begin{equation}
\tau \sim \exp{\frac{B}{\sqrt{|g - g_c|}}}
\label{eq:BKT_scaling}
\end{equation}
The agreement with Eq.~\eqref{eq:BKT_scaling} provides a compelling dynamical proof of the BKT nature of the transition, yielding a critical coupling $g_c$ that deviates by only $10^{-3}$ from the value obtained through equilibrium analysis \cite{signature}. This consistency confirms that the environmental memory effectively mediates the transition even in the absence of spatial dimensionality, establishing the relaxation time as an independent witness of the universality class. Complete relaxation profiles and fitting procedures are reported in the End Matter.

\textit{Kibble-Zurek Mechanism.} We investigate the non-equilibrium dynamics across a quantum critical point (QCP) by considering a Hamiltonian of the form $H(g)=H_0+gH_1$, which undergoes an equilibrium phase transition at $g=g_c$. The dynamics are driven by a linear quench protocol $g(t) = \frac{g_f}{t_Q} t$ for $0 < t < t_Q$, where $g_f > g_c$ and $t_Q$ denotes the quench duration. The temporal evolution is governed by the Schrödinger equation $i \partial_t |\psi(t)\rangle = H(t) |\psi(t)\rangle$. For systems exhibiting a second-order QPT, the critical slowing down is characterized by the divergence of the relaxation time in the vicinity of the QCP:
\begin{equation}
    \tau(g) \approx \tau_0|g-g_c|^{-z \nu},
\end{equation}
where $z$ and $\nu$ are the dynamical and correlation length critical exponents, respectively. This divergence leads to the breakdown of the adiabatic evolution near the QCP, leading to the population of excited instantaneous eigenstates of $H(t)$. 

According to the adiabatic-impulse approximation, the dynamics are divided into an adiabatic regime and an impulse regime where the state remains effectively frozen. For a linear quench, we define the residual time to the transition as $t_r(g)=\frac{t_Q}{g_f}|g-g_c|$. The crossover occurs at the freeze-out time $t_f$, satisfying the condition $\tau(g(t_f)) \approx t_r(g(t_f))$, which yields:
\begin{equation}
    t_f = A t_Q^{\frac{\nu z}{1+ \nu z}}.
\end{equation}
The KZM predicts that the correlation length freezes at $\bar{\xi} \propto t_f^{1/z}$. Consequently, the excitation probability $P_{\mathrm{exc}}(t)=1-|\langle\psi(t)|\psi_{GS}(t)\rangle|^2$ and the residual energy $E_r(t)=\langle\psi(t)|H|\psi(t)\rangle - E_{GS}(t)$ are expected to follow universal scaling relations:
\begin{align}
    P_{\mathrm{exc}} \propto t_Q^{-\frac{d \nu}{\nu z +1}} \sim t_f^{-d/z}, \\
    E_r \propto t_Q^{-\frac{(d + z) \nu }{\nu z + 1}} \sim t_f^{-d/z-1}.
\end{align}
These power laws, derived via adiabatic perturbation theory \cite{DeGrandi1,DeGrandi2,DeGrandibookr}, undergo significant modifications in the BKT universality class, where the freeze-out condition is governed by an essential singularity:
\begin{equation}
    A e^{\frac{B}{\sqrt{|g-g_c|}}} = \frac{t_Q}{g_f}|g-g_c|.
    \label{frezzeBKT}
\end{equation}
This transcendental equation can be solved exactly in terms of the Lambert $W$ function \cite{DziarmagaBKT} 
\begin{equation}
    t_f = \frac{t_Q}{g_f} \left(\frac{B}{2W\left(\frac{B}{2}\sqrt{\frac{t_Q}{Ag_f}}\right)}\right)^2.
    \label{freeze-analytic}
\end{equation}
While this implies that $t_f$ (and thus the observables) does not adhere to a simple power law of $t_Q$, the universal KZ scaling remains a robust witness of the transition when evaluated at the calculated freeze-out time.

In the zero-dimensional limit ($d=0$) relevant to our case study, the absence of spatial structure makes the conventional dynamical exponent $z$ ill-defined. Nevertheless, the emergence of KZ scaling is not restricted to spatially extended systems; robust evidence of universal non-equilibrium dynamics has been established for zero-dimensional systems, provided the quench terminates exactly at criticality \cite{d0_1}. We demonstrate that the observed scaling of the residual energy is consistent with an effective two-level mapping, where non-adiabatic excitations are generated primarily within the spin degrees of freedom \cite{Damski2005}. In this Landau-Zener-type framework, the excitation energy is predominantly governed by the closing of the renormalized spin gap $\Delta_{eff}$:
\begin{equation}
    E_{\mathrm{exc}} \approx P_{\mathrm{exc}} \Delta_{eff}.
\end{equation}
This phenomenological approach is corroborated by observations in $d=0$ systems where symmetric ramps (terminating far from the QCP) exhibit negligible residual energy, as the contribution from the gap closure is effectively recovered \cite{d0_2,d0_3}. Crucially, the relevant energy scale is the spin gap rather than the global gap between the ground and first excited states of the total system. Our results suggest that the bath continuum remains largely unexcited throughout the quench. This indicates that non-Markovian memory preserves the integrity of universal non-equilibrium scaling, revealing that $E_r$ evaluated at $t_f$ follows the predicted universal behavior even in the absence of spatial structure. 

\textit{Numerical results.} In order to investigate the dynamics after a quench, firstly we used Eq. \eqref{freeze-analytic} to calculate the freeze-out time for different quench durations (Fig 2a).

\begin{figure}[H]
    \centering
    \includegraphics[width=\linewidth]{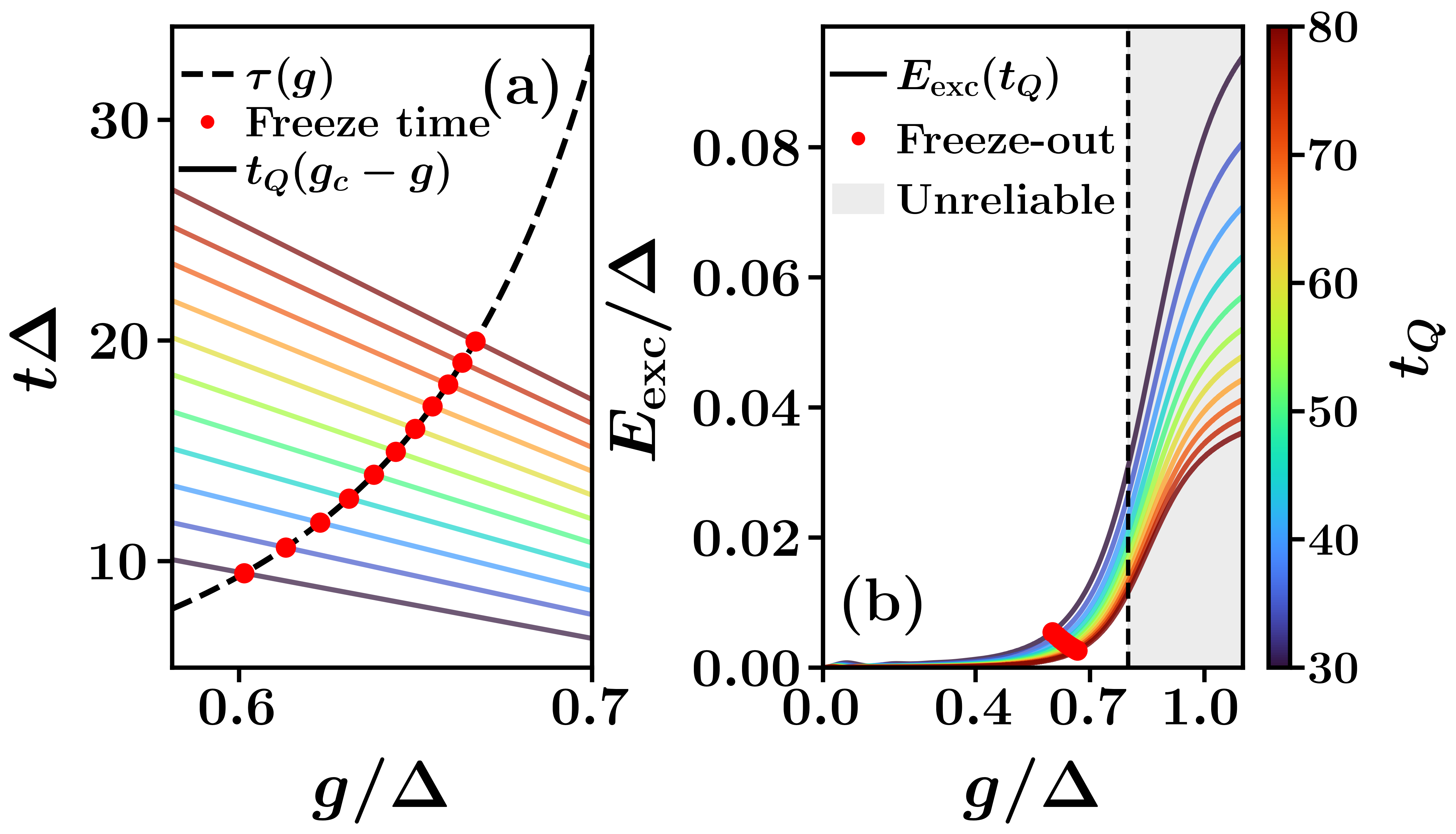}
    \caption{Freeze-out time $t_f$ (a) and excitation energy $E_{\mathrm{exc}}$ (b) vs $g/\Delta$. (a) Intersection of the BKT relaxation time $\tau(g)$ (dashed line) and the quench-induced residual time $t_r(g)$ (solid lines) for various $t_Q$: red dots denote the freeze-out time identification. (b) $E_{\mathrm{exc}}$ for different quench rates: red dots mark the breakdown of adiabaticity, with the shaded area indicating the numerically unreliable regime for $g > 0.8$.}
    \label{fig:freeze_out}
\end{figure}

Varying the quench velocity results in a shift of the freeze-out time, marking Tthe breakthrough of adiabaticity as illustrated in Figure 2(b). The system follows the instantaneous ground state until $t_{\mathrm{f}}$, beyond which non-adiabatic excitations are generated.  By evaluating the energy at this threshold, we observe that the excitation energy obeys a power-law scaling $E_{\mathrm{exc}} \propto t_{\mathrm{f}}^{-\mu}$, with an exponent $\mu = 0.992$ [Fig. 3(a)]. The breakdown of adiabaticity is also captured by the excitation probability $P_{\mathrm{exc}} = 1 - |\langle\psi_{GS}|\psi\rangle|^2$, which remains throughout a purely adiabatic evolution. Near the freeze-out time, the state acquires non-vanishing projections onto the excited eigenstates. We simulated the dynamics until $t_f$, calculated the instantaneous ground state with DMRG, and then determined $P_{\mathrm{exc}}$. Figure 3(b) shows $P_{\mathrm{exc}}$ vs $t_f$ follows a power-law scaling with exponent $\mu = 1.07$.

\begin{figure}[htbp]
    \centering
    \begin{minipage}{0.49\linewidth}
        \centering
        \includegraphics[width=\linewidth]{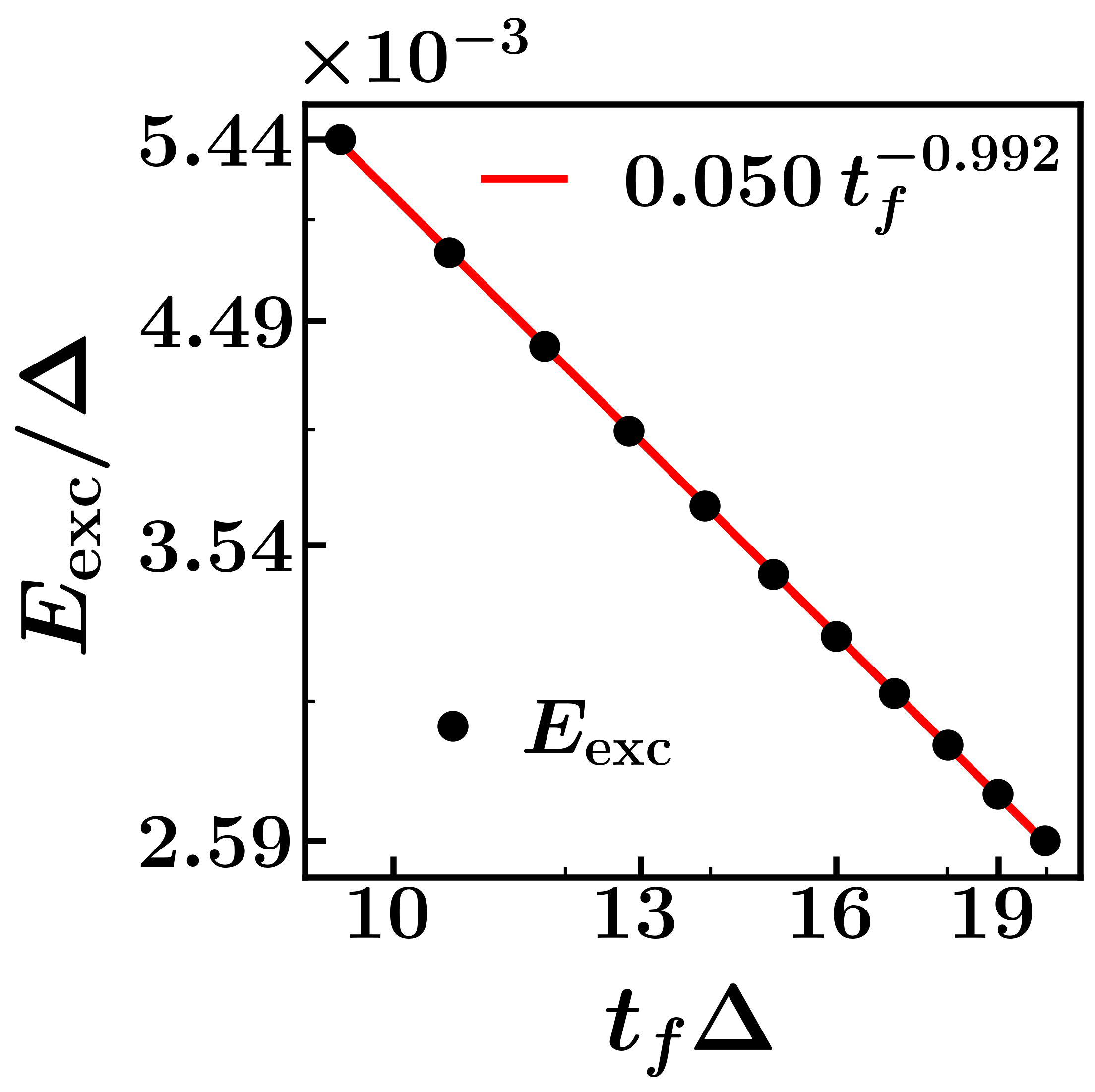}\\
        (a)
    \end{minipage}\hfill
    \begin{minipage}{0.49\linewidth}
        \centering
        \includegraphics[width=\linewidth]{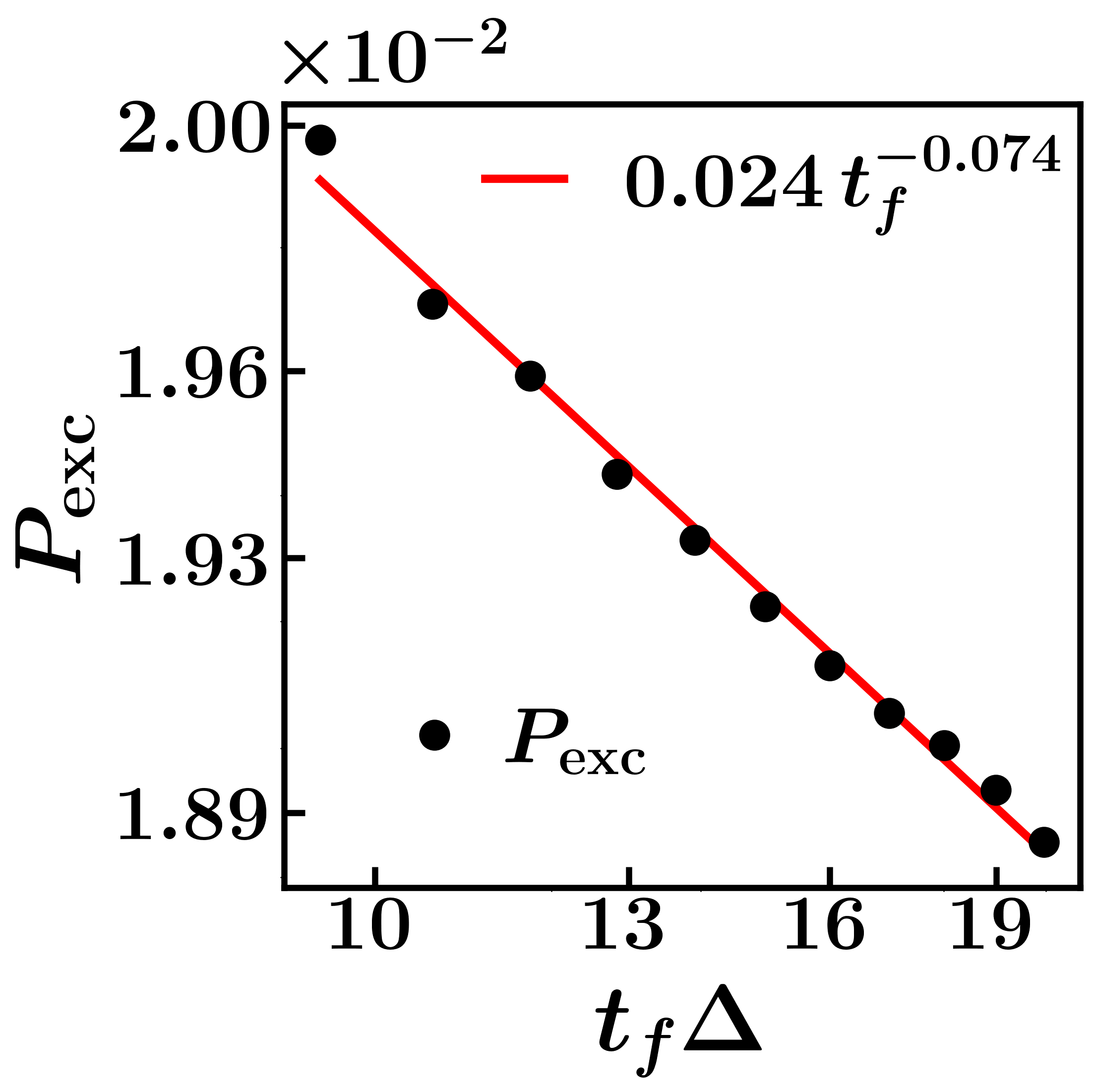}\\
        (b)
    \end{minipage}
    \caption{Excitation energy $E_{\mathrm{exc}}/\Delta$ (a) and excitation probability $P_{\mathrm{exc}}$ (b) vs $t_f\Delta$. (a) Power-law scaling of the residual energy: black dots denote numerical data, solid red line indicates the algebraic fit $\propto t_f^{-0.992}$. (b) Excitation probability scaling: solid red line marks the corresponding fit $\propto t_f^{-0.074}$.}
    \label{fig:energy_scaling}
\end{figure}

\begin{figure}[htbp]
    \centering
    \begin{minipage}{0.49\linewidth}
        \centering
        \includegraphics[width=\linewidth]{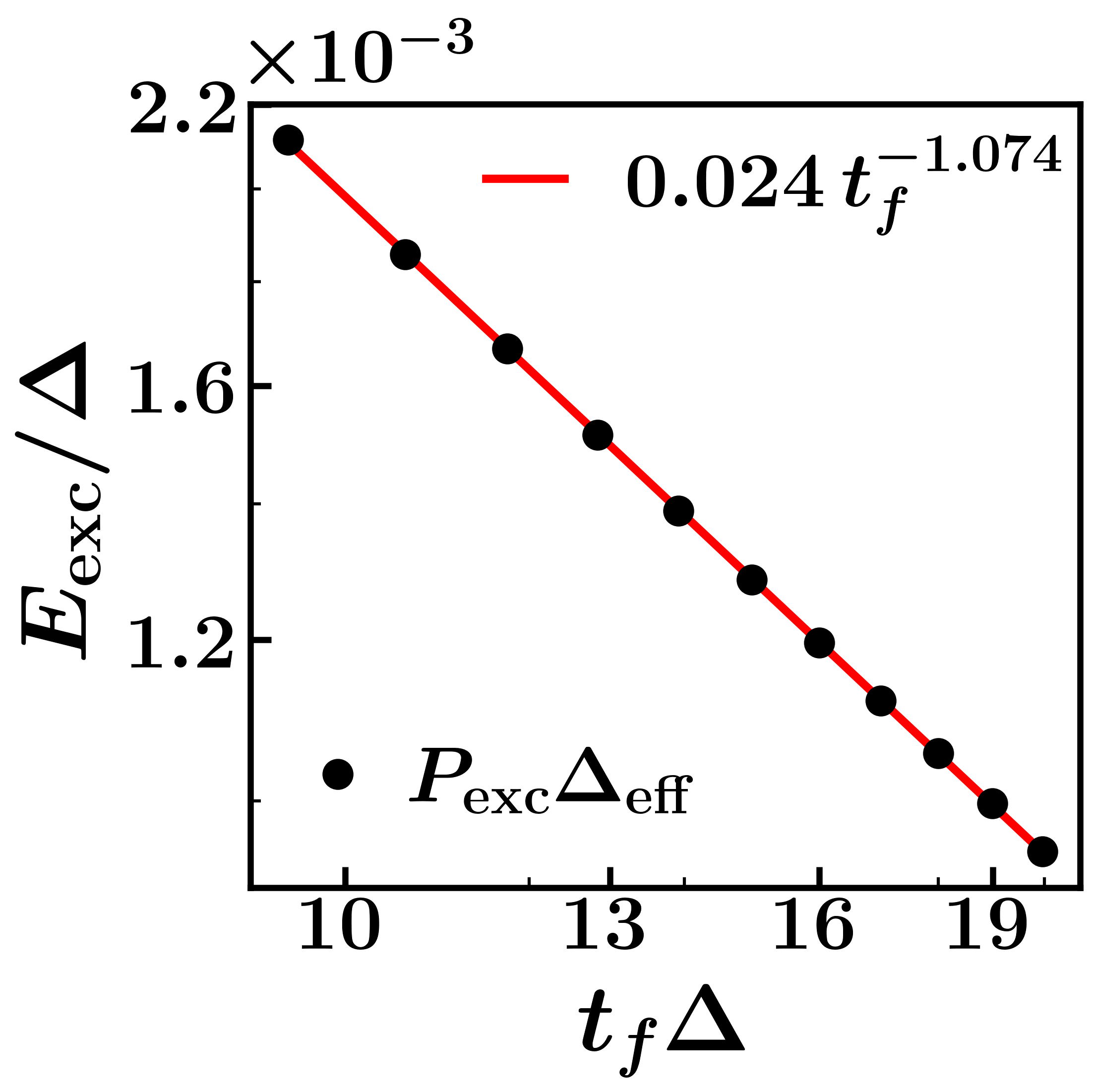}\\
        (a)
    \end{minipage}\hfill
    \begin{minipage}{0.49\linewidth}
        \centering
        \includegraphics[width=\linewidth]{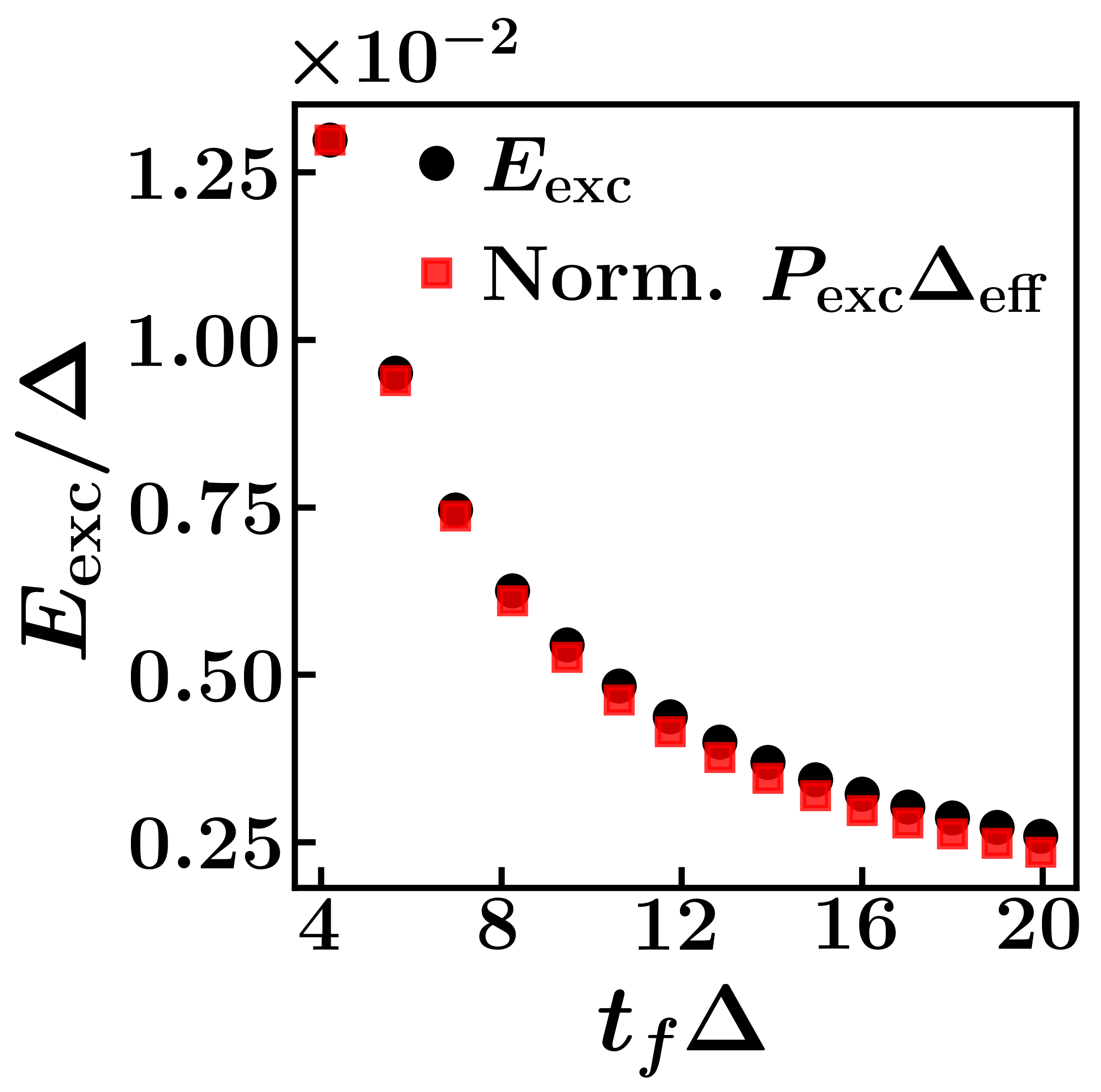}\\
        (b)
    \end{minipage}
    \caption{Estimated excitation energy $E_{\mathrm{exc}}/\Delta$ (a) and scaling collapse (b) vs $t_f\Delta$. (a) Estimated energy $P_{\mathrm{exc}}\Delta_{\mathrm{eff}}$ (dots) and power-law fit $\propto t_f^{-1.074}$ (solid line). (b) Measured $E_{\mathrm{exc}}$ (black dots) and $P_{\mathrm{exc}}\Delta_{\mathrm{eff}}$ rescaled to the first data point (red squares); the overlap confirms that both quantities share the same scaling behavior, differing only by a multiplicative factor.}
    \label{fig:comparison}
\end{figure}

The excitation energy can be estimated as $E_{\mathrm{exc}} \approx \Delta_{\text{eff}} P_{\mathrm{exc}}$, where the gap $\Delta_{\text{eff}}  = C \tau^{-1}$. Assuming $C=1$, Fig. 4(a) shows that the estimated $E_{\mathrm{exc}}$ follows a power-law scaling with an exponent $\mu = 1.074$, in close agreement with the value extracted from the measured energy. By rescaling the estimated energy to the measured value at the first data point [Fig. \ref{fig:comparison}], we find a remarkable overlap across the entire range. This result further supports the hypothesis that non-adiabaticity is dominated by transitions to the state separated by the effective gap $\Delta_{\text{eff}}$. 

For faster quenches (small $t_{\mathrm{f}}$), a systematic deviation appears where $E_{\mathrm{exc}}$ exceeds the two-level estimate. This discrepancy indicates that, in the rapid quench regime, the excitation process becomes more complex, potentially involving contributions beyond the simplest two-state mapping. Crucially, the fact that the global power-law scaling remains intact despite these deviations demonstrates that non-Markovian memory preserves the integrity of universal scaling even when the underlying excitation dynamics become non-trivial. 

\textit{Conclusions.} We have demonstrated the emergence of Kibble–Zurek scaling in the open quantum Rabi model subject to a non-Markovian Ohmic bath. By analyzing both relaxation dynamics and finite-time quenches, we provided dynamical evidence of the Berezinskii–Kosterlitz–Thouless transition induced by the environment and showed that the excitation energy exhibits universal power-law scaling at the freeze-out time. In contrast to Markovian dissipation, which typically competes with adiabatic dynamics and suppresses universal scaling in the slow-quench regime, we find that non-Markovian memory effects do not destroy the scaling behavior. Rather than acting as extrinsic noise, the environment defines the critical properties that govern the non-equilibrium response.
Our results establish the Kibble–Zurek mechanism as a robust dynamical probe of criticality even in zero-dimensional open quantum systems, and demonstrate that environmentally induced quantum phase transitions can sustain universal non-equilibrium scaling.

\textit{Acknowledgements.} G.D.F. acknowledges financial support from PNRR MUR Project No. PE0000023-NQSTI. C.A.P. acknowledges funding from IQARO (Spin-orbitronic Quantum Bits in Reconfigurable 2DOxides) project of the European Union's Horizon Europe research and innovation programme under grant agreement n. 101115190.

\bibliography{references2}

\section*{End Matter: Relaxation Dynamics and Relaxation Time Estimation}

The relaxation time $\tau$ is estimated from the asymptotic decay of the longitudinal magnetization $\Sigma_z(t)$. Here, we provide a detailed analysis of the underlying temporal dynamics across different coupling regimes leading up to the critical point. As discussed in the main text, exact diagonalization of the bosonic sector maps the OQRM onto an effective spin-boson model with a structured bath. Crucially, the effective Ohmic spectral density behaves as $J_{\mathrm{eff}}(\omega) \approx (\alpha_{\mathrm{eff}}/2) \omega$ for $\omega \to 0$, characterized by an effective dissipation strength $\alpha_{\mathrm{eff}} = 4g^2\alpha/\omega_0^2$. The long-time dynamics is heavily dominated by this low-frequency behavior. Consequently, increasing the atom-field coupling $g$ directly enhances the effective coupling with the bath, ultimately driving the system toward the Berezinskii-Kosterlitz-Thouless (BKT) critical point.

\begin{figure}[H]
    \centering
    
    \begin{minipage}{0.48\columnwidth}
        \centering
        \includegraphics[width=\linewidth]{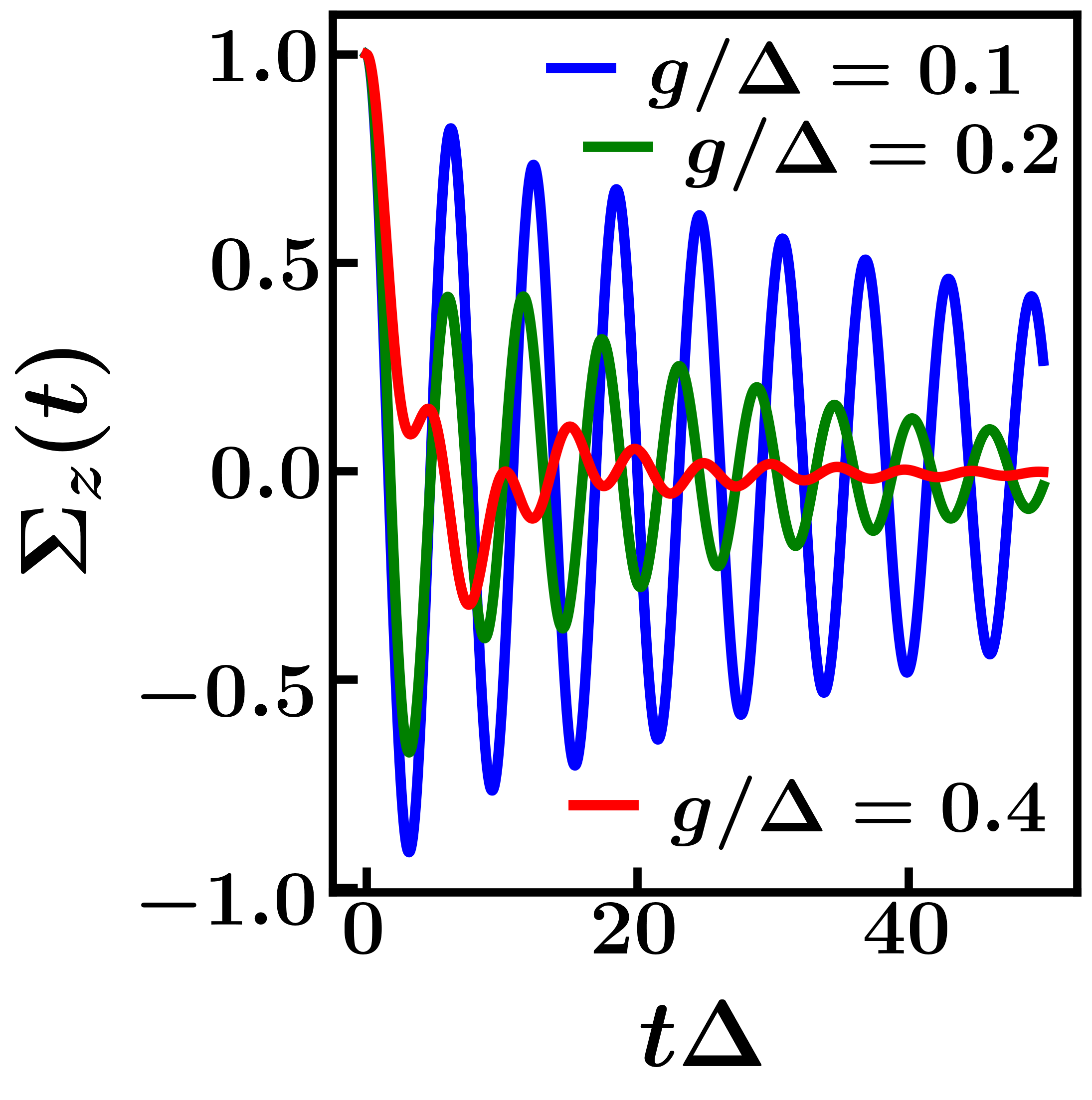}\\
        (a)
    \end{minipage}\hfill
    \begin{minipage}{0.48\columnwidth}
        \centering
        \includegraphics[width=\linewidth]{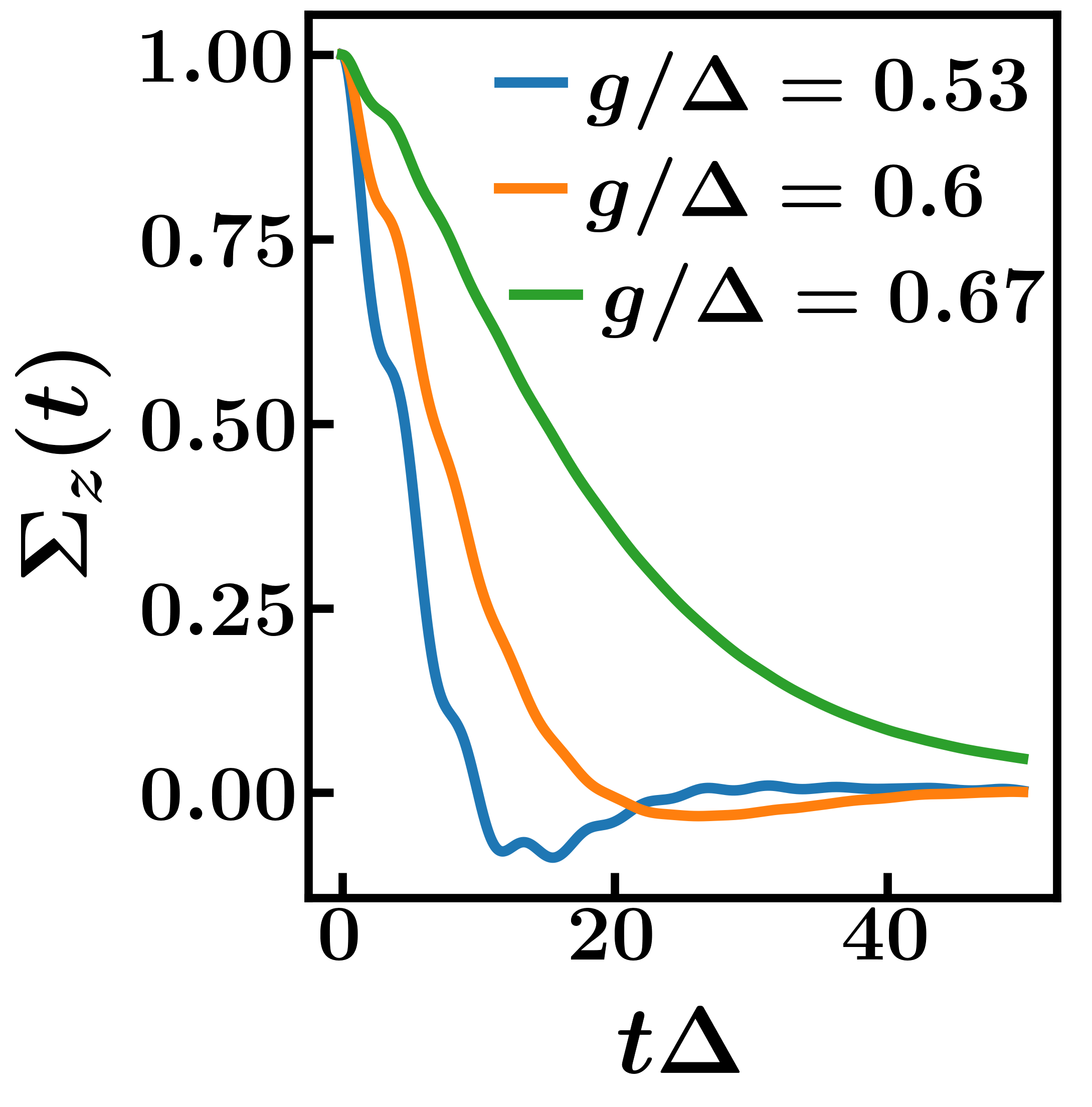}\\
        (b)
    \end{minipage}
    
    \vspace{0.4cm}
    
    \begin{minipage}{0.48\columnwidth}
        \centering
        \includegraphics[width=\linewidth]{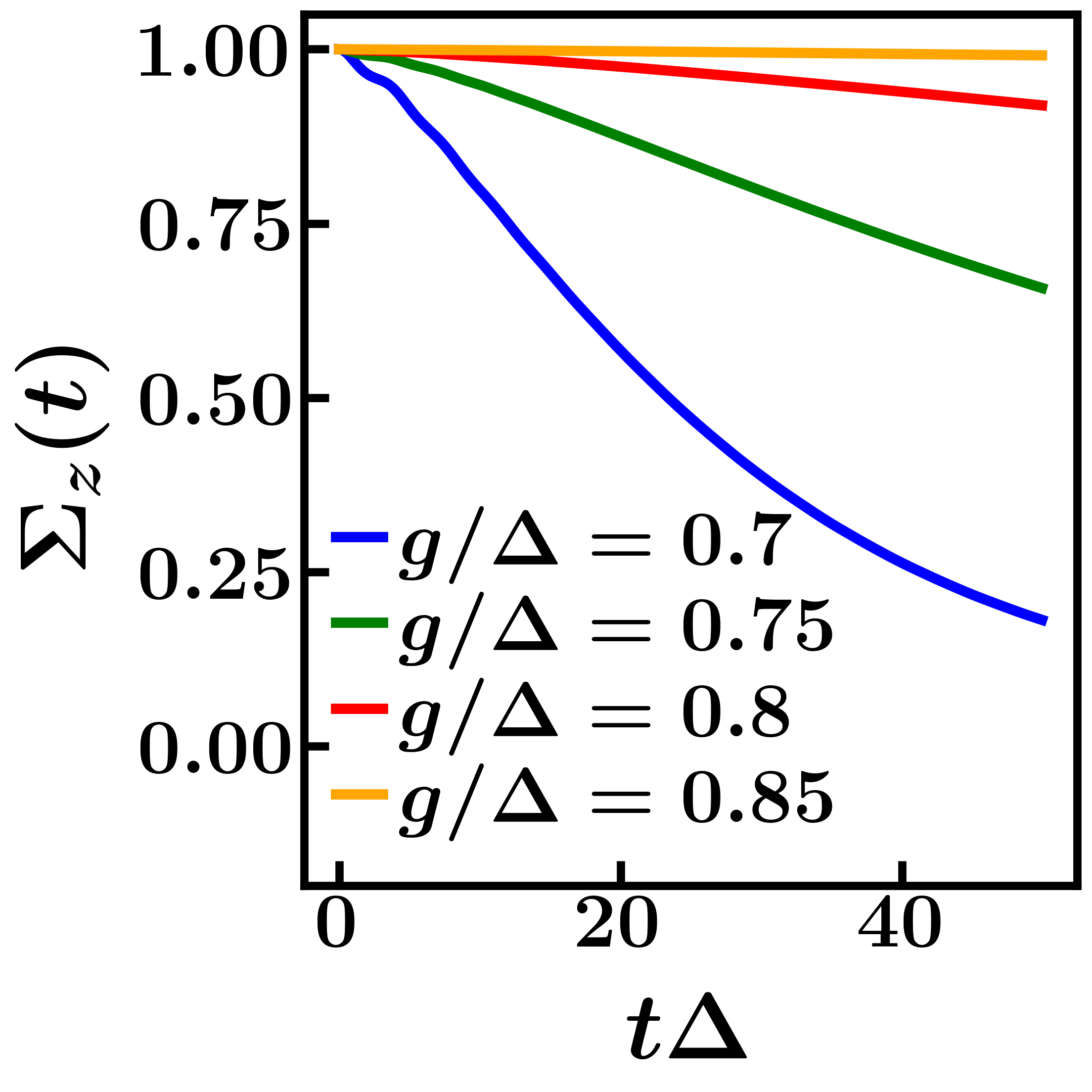}\\
        (c)
    \end{minipage}
    
    \caption{Relaxation dynamics of $\Sigma_z(t)$ in the OQRM. (a) Coherent oscillations at weak coupling. (b) Crossover to critically damped behavior near the Toulouse point. (c) Strictly overdamped regime approaching the quantum critical point. Couplings $g/\Delta$ are indicated in each panel.}
    \label{fig:relaxation_dynamics_panels}
\end{figure}

\begin{figure}[htbp]
    \centering
    \begin{minipage}{0.48\columnwidth}
        \centering
        \includegraphics[width=\linewidth]{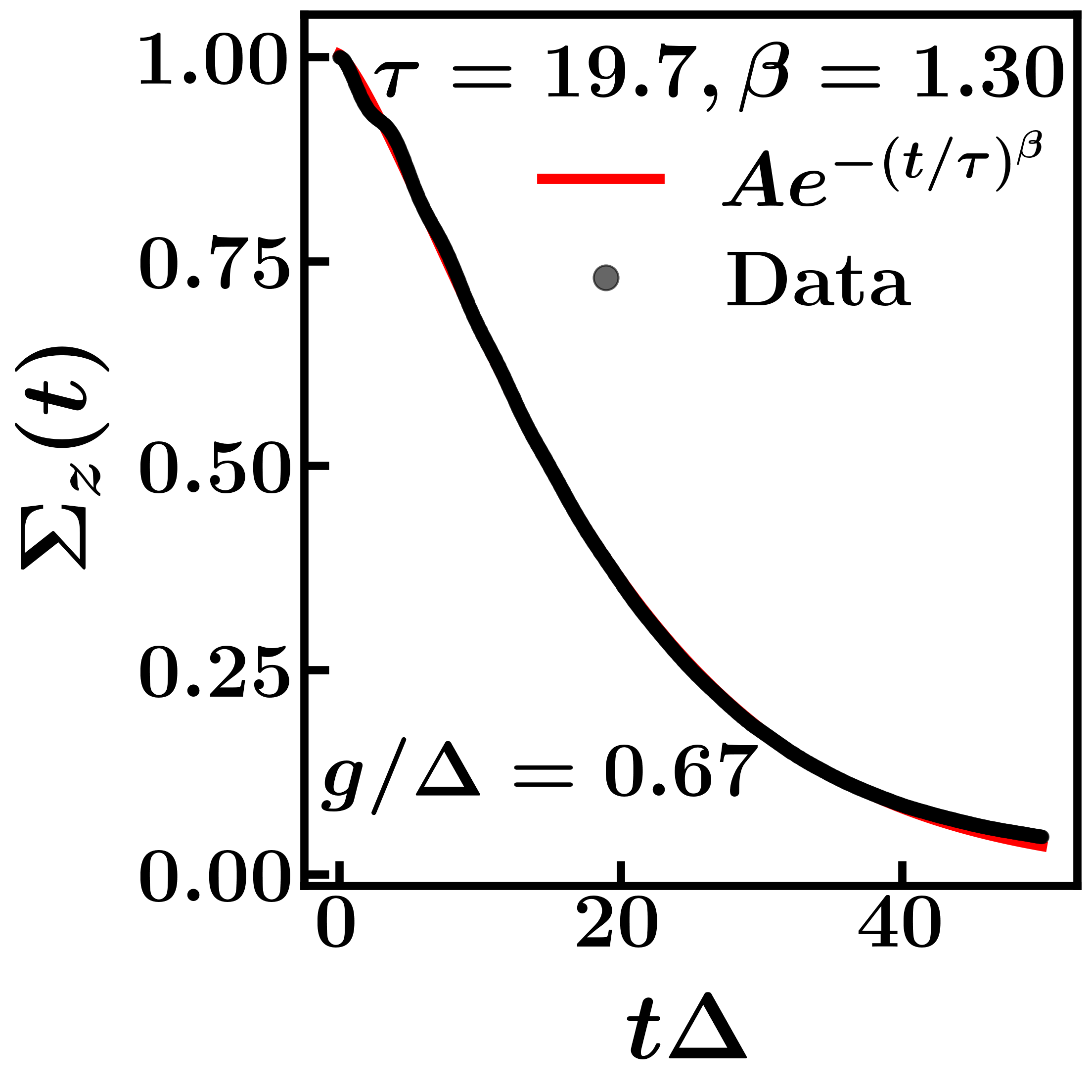}\\
        (a)
    \end{minipage}\hfill
    \begin{minipage}{0.48\columnwidth}
        \centering
        \includegraphics[width=\linewidth]{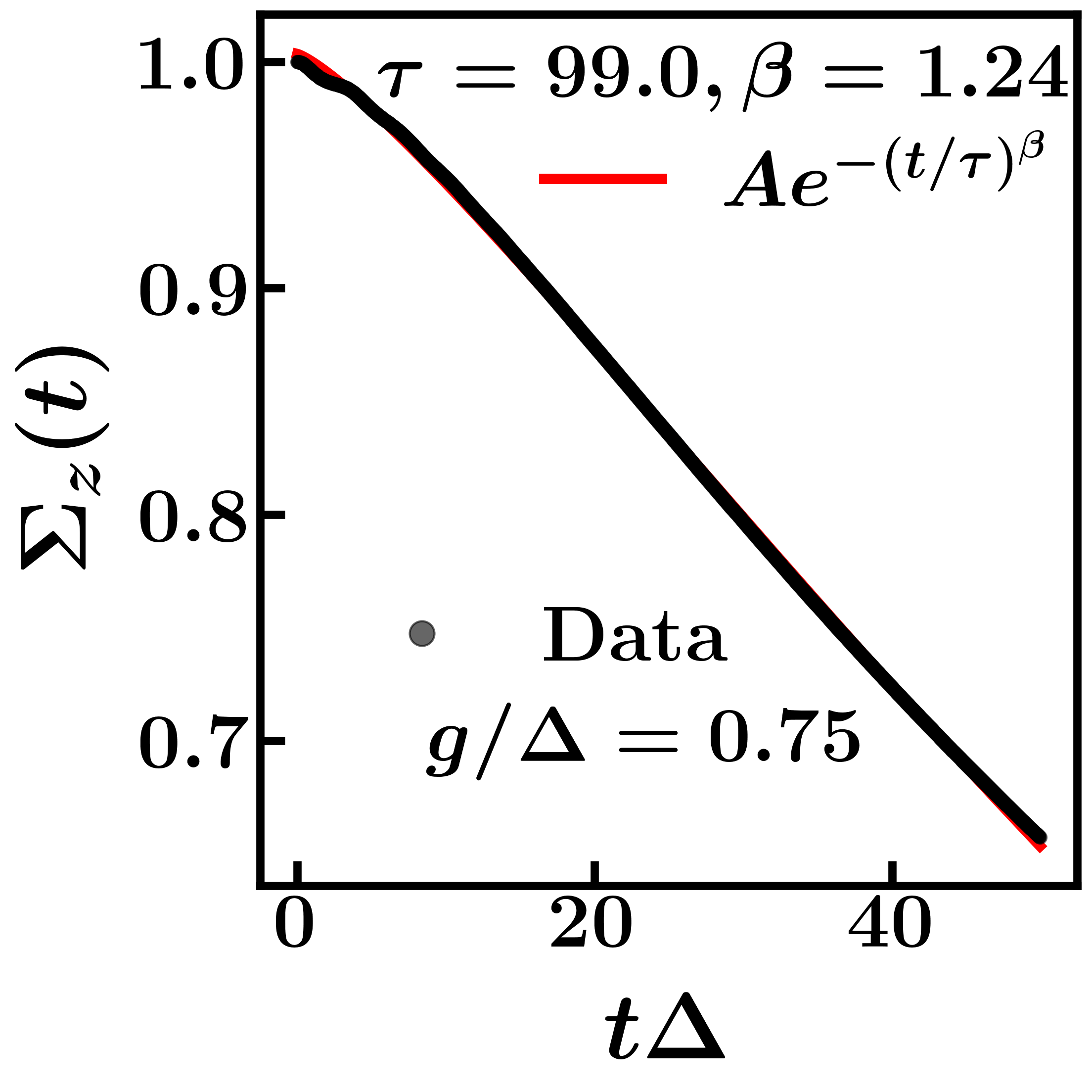}\\
        (b)
    \end{minipage}
    \caption{Estimation of the relaxation time $\tau$ in the overdamped regime. The numerical data for $\Sigma_z(t)$ (dots) are fitted against a stretched exponential function (solid lines). Representative plots are shown for (a) $g/\Delta = 0.67$ and (b) $g/\Delta = 0.75$.}
    \label{fig:relaxation_fit_details}
\end{figure}

This evolution is explicitly captured in Fig.~\ref{fig:relaxation_dynamics_panels}. In the weak coupling regime ($g/\Delta \le 0.4$), the dynamics is characterized by coherent Rabi oscillations undergoing standard environmentally induced damping [Fig.~\ref{fig:relaxation_dynamics_panels}(a)]. As the coupling increases toward $g/\Delta \sim 0.6$, the system relaxes without completing a single full oscillation. This crossover signature corresponds to the Toulouse point of the effective spin-boson model [Fig.~\ref{fig:relaxation_dynamics_panels}(b)]. Beyond the Toulouse point, the dynamics enters a strictly overdamped regime [Fig.~\ref{fig:relaxation_dynamics_panels}(c)]. The coherent oscillations are completely suppressed, and the relaxation process becomes a purely overdamped decay. As $g$ approaches the critical value $g_c \simeq 0.9165$, the bath-induced dissipation effectively inhibits the tunneling between the spin states, leading to a localization of the system \cite{lehur-entangl,LeHur2008}. In this limit, the dynamics effectively freezes, reflecting the divergence of the relaxation time at the phase transition.
\\
To estimate the relaxation time $\tau$ used for the critical scaling, we focus on the strictly overdamped regime. Here, the relaxation deviates from a simple exponential and is instead accurately described by a stretched exponential form, $\Sigma_z(t) \propto \exp[-(t/\tau)^\beta]$. By fitting the numerical data directly to this model, we consistently obtain a stretching exponent $\beta \approx 1.2$, which captures the complex multi-scale relaxation induced by the non-Markovian bath. As shown in Fig.~\ref{fig:relaxation_fit_details} for representative couplings ($g/\Delta = 0.67$ and $0.75$), this stretched exponential form perfectly fits the long-time dynamics, allowing for a highly reliable estimation of $\tau$. Crucially, the agreement between our numerical results and the expected phenomenological behavior of the spin-boson model provides a robust validation of the TDVP simulations in capturing the dissipative dynamics of the OQRM.

\end{document}